\documentstyle[12pt]{article}
\date{}
\begin{document}
\vspace*{1cm}
\hspace{3cm}
\begin{flushright}
Tbilisi Institute of Physics~~~~~~~~~~~~~~~~~~~~~~~~HE--7/94/E\\
October 1994\\
\end{flushright}
\medskip
\begin{center}
{\Large \bf On dynamics of pseudorapidity fluctuations\\
in central C--Cu  collisions at 4.5~A GeV/c}\\
\vspace{0.5cm}
{\bf E.K. Sarkisyan\footnote{E-mail addresses:
saek, gelovani, goga@physics.iberiapac.ge}$^,$\footnote{Present address:
School of Physics and Astronomy, Tel Aviv University, Tel Aviv, Israel;
E-mail address: edward@lep1.tau.ac.il}, L.K. Gelovani$^1$,
G.L. Gogiberidze$^1$}\\
\smallskip
{\small \it Institute
of Physics, Tamarashvili street 6, GE-380077 Tbilisi,  Georgia}\\
\medskip
{\bf and}\\
\medskip
{\bf G.G. Taran}\\
\smallskip
{\small \it
P.N.~Lebedev Physical Institute,
Leninsky prospect 53, RU-117924 Moscow,\\ Russian Federation}
\end{center}
\vspace{0.5cm}

\begin{abstract}

Results on dynamical fluctuations of charged particles in the pseudorapidity
space of central C--Cu interactions at 4.5 $A$ GeV/$c$ are performed in the
transformed variables and using higher order scaled factorial moments
modifyied to remove the bias of infinite statistics in the normalization.
The intermittency behavior is found up to eighth order of the moments
increasing with the order and leading to the pronounced multifractality. Two
differed intermittent-like rises are obtained, one indicating an occurrence of
the non-thermal phase transition, and no critical behavior is found to be
reached in another case. The observations may be treated to show different
regimes of particle production during the cascade. Comparison with some
conventional model approximations notes the multiparticle character of the
fluctuations. The results presented can be effective in sense of sensitivity
of intermittency to the hadronization phase.

\end{abstract}

\vspace{0.6cm}

\centerline {Submitted to {\sl Physics Letters B}}
\newpage
\footnotesize
\section {Introduction}

At present intermittency \cite{bp1} seems to be a well-founded fact
observed in high-energy multiparticle production \cite{dw,ch}. This
phenomenon, expressing as the power-law behavior,

\begin{equation}
\langle F_q\rangle \, \propto M^{\varphi_q},
\qquad 0<\varphi_q\leq q-1,
\label{fi}
\end{equation}
of the $q$-order normalized scaled factorial moments (SFM),

\begin{equation}
\langle F_q\rangle =
\frac{1}{M}
\sum_{m=1}^{M}
\frac{\langle n_m^{[q]}\rangle }
{\langle n_m \rangle ^q}
\label{f}
\end{equation}
(``vertical'' analysis \cite{dw}--\cite{pijmp}), identifies
dynamical fluctuations. Here $n_m^{[q]}$ is the $q$th power factorial
multinomial, $ n_{m}\, (n_m-1)\cdots (n_m-q+1) $, with multiplicity $n_m$
in the $m$th bin from $M$ ones, into which the space of the particles
produced are divided.  Average is taken over all events.

The intermittency indexes, $\varphi_q$, obtained via (\ref{fi}) are shown
\cite{lb,s,h} to be related to the fractal codimensions $d_q$
\cite{pv,d},

\begin{equation}
\varphi_q=(q-1)\, d_q \; ,
\label{d}
\end{equation}
and correspondingly reflects the inner structure of fluctuations
representing monofractal patterns with the unique $d_q$, or multifractal
ones with the hierarchy $d_q>d_p \;, \; q>p$.

Applied to multihadron production processes, formation of such geometrical
structures are pointed out \cite{bqm,pl} to be a manifestation of one of
two possible mechanisms. The first one, leading to the $q$-dependence of
$d_q$, is that one connects \cite{pijmp} with self-similar cascade, rather
a ``non-thermal'' (non-equilibrium) phase transition during the cascade
than particle creation within one phase, e.g.
hadronic \cite{ow}.  The second scenario assumed
at monofractality, is associated with thermal transition \cite{bqm,dw},
e.g.  from a quark-gluon plasma expected to be reached in central
collisions of (ultra)relativistic nuclei \cite{hj}.

So study of intermittency decodes through fractality the geometrical and
then thermodynamical features of high-energy multiparticle production. Note
that the random cascading models being widely used as suitable to describe
such processes (e.g., negative binomial distribution) \cite{k},  lead
in the most general form to the scaling (\ref{fi}) \cite{bp1} , though
they seem to be problematic to reproduce intermittency observed
\cite{fos,dm,du}. In general, the observations are still far from the
qualitative explanation using existed particle-production codes, and
origin of intermittency/fractality is a matter of debates \cite{dw}.

Meanwhile, intermittency turns out to be very sensitive to the hadronic
phase \cite{dw,pijmp}. Since the interpretation is mostly done in the
approach based on the parton cascade models, it is rather difficult to
explain  the fact that partonic local fluctuations survive the
hadronization process \cite{ow,hadph}. In this context it is noticeable
that ``universality'' of multihadron production in different type of
reactions (from lepton-hadron to nucleus-nucleus) \cite{fu} can be just a
reflection of hadronization dynamics properties \cite{haddyn}.  Thus,
search for dynamical fluctuations and its possible treatment in ``soft''
process terms attracts considerable interest in reactions at
intermediate energies.

Important features of intermittency systematics behavior
found in high-energy nuclear interactions
\cite{s1}-\cite{sh} are also
already found to be manifested at relativistic energies. Besides purely
hadronic model calculations showing intermittent structure due to the
random cascade \cite{li1}, enhancement of intermittency with beam energy
decrease \cite{li2} and weakening of the effect with increase complexity
of reaction \cite{inc,yaf3} are obtained. Recently, processes with low
average multiplicity was discussed \cite{dj,du} to dominate in the
fractional SFM analysis.

\section{Experimental procedure}

The present paper deals with the study of the data
obtained after processing the pictures of the 2m Streamer Chamber
SKM-200 \cite{a1} equipped with a copper  target.
The chamber was installed in a 0.8 T magnetic field and it was
exposed to the 4.5~$A$ GeV$/c$ $^{12}$C beam at the JINR Synchrophasotron
(Dubna).  In data taking, the central collision trigger was used:  the
chamber was started if there were no charged particles with momenta
larger than 3 GeV$/c$ in a forward cone of 2.4$^{\circ }$.
Details of the set-up design and data reduction procedure are described
elsewhere \cite{a1,a2}.  Systematic errors related to the trigger effects,
low energy pion and proton detection, the admixture of electrons etc. have
been considered in detail earlier \cite{a3} and the total contribution does
not exceed  $3\%$.

The scanning and handling of the film data were carried out on
special scanning
tables of the Lebedev Physical Institute (Moscow), using the method elaborated
in ref. \cite{t}. The average measurement error in the momentum
$\langle \varepsilon_p/p\rangle $ was about $12\% $, and that in the
polar angle measurements was $\langle \varepsilon_{\vartheta}\rangle \simeq
2^{\circ }$. To search for dynamical fluctuations  the charged
particles in the pseudorapidity ($\eta =
- \ln \tan (\vartheta/2)$)
region $\Delta\eta=0.2-3.0$ (in the
target rest-frame) were used, in which the angular measurement accuracy was
not larger than 0.1 in the $\eta$-units. The samples of 305 C--Cu events,
which meet the above centrality criterion, have been selected with the
average multiplicity of $27.2\pm 0.8$ in the $\Delta \eta$ under
consideration.

Earlier we have already analyzed fluctuations in the data presented and
existence of non-statistical (dynamical) fluctuations was shown
\cite{yaf3,pl2,pl3} within intermittency approach and
using multifractal analysis \cite{h,dw}. It is noteworthy that only
applying the methods with statistical background suppression allowed dynamical
nature of the fluctuations to be manifested, while the comparison between
data and completely uncorrelated particle-production simulation fails to do
that \cite{yaf1-2}. It is significant in sense of the above discussed
``hadronic origin'' of the intermittency that analogous observation have
been done at ultra-relativistic energies \cite{1n} (see also ref.
\cite{1l}).

The study \cite{pl2,pl3} also showed multifractality of the particle spectra
along
with multiparticle character of the fluctuations  being
important to choice  of the real particle-production model,
as discussed. Multifractality
have been also observed at ultra-relativistic heavy-ion  collisions
but obliged mainly to the two-particle correlations \cite{dw,ch}.
Note that
multiparticle contribution to the very-short-range correlations are
directly observed in our data using the method of factorial cumulants
\cite{prep} as found in hadronic interactions recently \cite{22z}.

Before presenting the results, two important technical remarks should be
made.  First of all, in the previous investigations we have used
``horizontal''  analysis of the SFM
taking into account the non-flat shape of the
one-particle pseudorapidity distribution $\rho(\eta)$  to minimize the
difference from ``vertical'' normalization (\ref{f}) \cite{ch,pwh}. Note
that, though the ``corrected'' SFM were widely applied  to analysis (in
particular, at relatively small multiplicities) it implies the
fluctuations to be bin-independent (the sum of the quotient in (\ref{f})
transforms into the fraction of independent sums, $\sum_{m=1}^{M}\langle
n_m^{[q]}\rangle/\sum_{m=1}^{M}\langle n_m \rangle ^q$) that is,
generally, a non-trivial  assumption \cite{pwh,pijmp}. In our data the
slopes of the vertically-normalized moments are significantly greater
the corrected ones at $\delta\eta<0.5$ \cite{prep}, while they
coincides in collisions of ultra-relativistic ions \cite{1n}.

To overcome the problem and, moreover, to compare the
results observed in different experiments a new transformed variable,

\begin{equation}
\stackrel{\sim }{\eta}(\eta)=
\frac{
\int_{\eta_{min}}^{\eta} \rho(\eta')d\eta'}
{\int_{\eta_{min}}^{\eta_{max}} \rho(\eta')d\eta'
} \; ,
\label{nv}
\end{equation}
have been introduced \cite{nv,o}, so that $\stackrel{\sim }{\eta}$ is
uniformly distributed
in the [0;1] interval
($\rho(\stackrel{\sim }{\eta}) \approx {\rm const}.$).
Due to the scale properties of the variables (\ref{nv}), the one-particle
spectrum stretches in its central region (not significantly changing at the
$\Delta \eta $-edges), eliminating losses from bin-splitting and, thus,
allowing to observe higher-order moments.

Another remark regards to the biased estimator of the SFM normalization,
being the $q$-particle density function for uncorrelated production of
particles in assumption of infinite statistics \cite{bp1,pwh,pijmp}. This
sensitively influence the scaling law (\ref{fi}) for small bins.
Note that a flattening of the moments for  $M\ge M_0$ is expected to be
a reflection of the attainment of the correlation length \cite{bp1}.
Really, measured SFM go to zero as the bin size aspires to the
experimental resolution \cite{1n}, or much before because of the
statistics limitations (``empty bin effect'' \cite{eb,eb2}).

Recently, the modification of the method of the SFM have been proposed
\cite{ks} to remove the bias in the normalization, and then the bias-free
moments are  defined to be , e.g. at ``vertical'' analysis,

\begin{equation}
\langle F_q\rangle =
\frac{{\cal N}^q}{M}
\sum_{m=1}^{M}
\frac{\langle n_m^{[q]}\rangle }
{N_m^{[q]}}\; ,
\label{fb}
\end{equation}
where $N_m$ is the number of particles in the $m$th bin in all $\cal N$
events, $N_m= \sum_{j=1}^{\cal N}(n_m)_j$.
The property of the ``vertical'' and ``horizontal''
analysis to give the same results if the scaled
variables (\ref{nv}) are used ($n_m\approx\langle
n\rangle/M$) seems to be also valid for the definition
(\ref{fb}).

\section{Results and discussion}

In fig.1 we show the log-log plots of the modified SFM (\ref{fb}) vs.
number of bins of the space of ``pseudorapidity'' (\ref{nv}).
Though by illustration reasons  the dependencies
are given  only for
some  orders, viz. $q=3,5,6,8$, they reflect the common
peculiarities  to be noticed. From the plots and the values of
$\varphi_q$ (see table) one can conclude that,
besides considerable intermittent behavior, pronounced  up to higher
orders, there are the different power-like dependences in different
intervals.   Since at low orders ($q=2,3$) this expresses as two
sharply differed slopes: the values at $M\le 22$ are about seven times less
than ones at $M\ge 23$, this effect grows weak at $q=4$ and $5$. It is
visible also that as the order increase an additional intermittent
structure manifests at the range of small bins, but doesn't in fact
survive when five and more particles are required to fill the bin and
effect of small statistics becomes sizeable.

However, regardless to the irregularities in the $M$-dependence of the
SFM, $\varphi _q$ increase with the order up to higher moments. Moreover, in
the $M$-interval of [7;17] and especially at $10\le M\le 17$ strong
intermittency is seen.

It should be noted that the irregular behavior of the SFM was also
observed in our previous investigation \cite{yaf3}, but using of the
``ordinary'' pseudorapidity variable strongly limited the order of the
moment.  Transforming $\eta $-spectra into the uniform ones made this
effect more pronounced and allowed to consider the SFM of high orders,
where a new sub-structure is revealed.  Note    that strongly non-linear
increasing of the log-log plots as presented have been observed in
different interactions, from lepton-hadron to heavy ion collisions
\cite{dw}, particularly at high-orders of the moments (e.g.
\cite{s1}) and/or at multidimensional analysis \cite{s2,1n,35}.

As discussed, intermittency index, or rather codimension (\ref{d})
dependence on $q$-order gives an information on the possible
particle-production scenario. In fig.2 we demonstrate $d_q$ as a function
of $q$ for different $M$-intervals,   within which a linear fit of the SFM
plots (fig.1) is valid; the corresponding $\varphi_q$-values are shown in
the table. From the behavior of the $d_q$ obtained one can definitely
conclude that very multifractal structure ($d_p > d_q\,, p>q$) of
multiparticle production is revealed, independent of the interval of $M$
considered; no evidence for monofractality, and then for a
second-order phase transition is seen.
Let us to mention that decrease of the $d_q$ at high order is, in our
opinion, connected with relatively small number of particles per event.
Meantime, no considerable difference are obtained for some
$M$-intervals, namely for  the couples of
$7\le M\le 17$, $10\le M\le 22$ and $4\le M\le 15$, $2\le M\le 22$.

The multifractality, as noted above, lends support to find the
condition for the non-thermal phase, not characterized by a
thermodynamical behavior \cite{pijmp}. As signal of the transition to this
phase the existence of a minimum of the function

\begin{equation}
\lambda_q = \frac{\varphi_q + 1}{q}
\label{l}
\end{equation}
at a certain ``critical'' value of $q=q_c$ is required.
In such a ``spin-glass'' phase the events are dominated by a few
clots in projected distributions, while the ``normal'' phase
is represented by the events with a bulk of peaks and holes \cite{bz,bps}.
The minimum of eq. (\ref{l}), as shown \cite{bz},  may be a manifestation
of the fact that these two different phase  mixed. Moreover, the phase
transition can accompany (or occur inside) the branching process
\cite{pl,pijmp}.

In fig.3 the $\lambda_q$ are displayed as function of the $q$-order for
different $M$-intervals with considerably distinguished $d_q$ (see fig.2). A
clear minimum at
$7\le M\le 17$ ($10\le M\le 22$) and $4\le M\le 15$ ($2\le M\le 22$) are
seen, whilst the fits at $2\le M\le  28$ do not exhibit such a feature. In
our opinion, this is very intrinsic finding.

Indeed, while sensitive behavior of the SFM in the region of $2\le M\le
22$ are observing right up to eighth order (fig.1 and the table) such a
sharp difference of the dependence of $\lambda_q$ may be a signal of a
non-trivial dynamical effect\footnote {
The contribution of the statistics limitations is estimated to be small
for these $M$-intervals.}.
Referring to the non-thermal phase transition interpretation \cite{bpnp}
one should conclude that fits at different $M$-intervals lead to different
cases of the intermittency:  ``weak'' intermittency for monotone function
(\ref{l}) (large $q_c$) and ``stronger'' intermittency when minimum is
reached at $q=4$ or $5$. The stronger intermittency is related to the
so-called ``peak transition'', i.e. transition at positive index $q_c$,
and the weak case contains the absence of the transition also.  In our
view, such an ambiguity   is related to the cascade mechanism description
that includes many steps irregardless to {\it the change of the regime of
particle production at different bin-averaging scales}. This is,
that apparently brings indetermination in treating of the intermittency
observations e.g.  in $e^+e^-$-annihilation \cite{bpnp}  and, meanwhile,
being manifesting in our data is an extra proof of hadronic nature of
intermittency observed \cite{hadph}.

Noteworthily, only using of higher-order moments makes clear the hint
to the non-thermal transition found earlier slightly   visible up to
fifth order \cite{s2,pl3,sh}.
Whether the minimum is taken place, the low-$p_t$ effect
in the intermittency manifested at hadronic interactions \cite{22z} leads
to a clear minimum confirming responsible of hadronization for
the effect of intermittency \cite{bqm}. The $p_t$ range for the stronger
intermittency in our case was observed to be of 0.35--0.45 GeV/$c$ like one
have been already found searching for the maximum fluctuations
\cite{yaf1-2}.  It should be stressed also that since the
minimum of the $\lambda_q$ (\ref{l}) corresponds \cite{dw} to zeros of the
fractal spectra, the system ``frozen'' at this point is no longer
self-averaging and introducing of new observables is needed \cite{bsz}.

The shown importance of the cascade approach in the multiparticle
production and the abovementioned underlying of these processes to
describe the hadronization attracts considerable interest to compare the
predictions with the observations. Let us to limit oneself what have been
earlier studied related to the SFM method, viz. negative binomial
distribution  (e.g. \cite{dm}), gaussian approximation
\cite{bp1} and scale-invariant mass-splitting branching model \cite{ow}.

If the negative binomial distribution (NBD) is valid then the SFM are
determined by the recurrence relation

\begin{equation}
\langle F_q \rangle =
(1 + \frac{1}{k})\,
(1 + \frac{2}{k})\cdots
(1 + \frac{q-1}{k})\, ,
\label{nbd}
\end{equation}
where $k$ is one of two NBD parameters should be independent of the
$\delta \eta$. The bin dependence of $k$ expected from the intermittent
rise at $q=2$ (see table) is a reflection of the instability of the NBD
\cite{F}. Besides, the NBD is a good fit of the data in the central
$\eta$-region only, and does not describe the ``tails'' of the
multiplicity spectra, for which it transforms into the
$\Gamma$-distribution being stable \cite{F}. In the previous ref.
\cite{pl3} we have shown that at the values of $\varphi_q$ closed to the
 experimentally obtained ones, the absolute values of the SFM approach
each other when the shape of the $\eta$-spectra is accounted for. From
this observation and the noted ``central property'' of the NBD  the
satisfactory agreement of the moments based on (\ref{nbd}) with the
experimentally measured ones in the ``flat'' spectra over the scaled
variable (\ref{nv}) have been predicted. This is a fact well seen now from
fig. 1 ($\chi^2/NDF$  for the $\langle F_q\rangle $ are less than one
standard deviation in all the $M$-regions), may be excluding the case of
$q=8$ (e.g. $\chi^2/NDF\approx 6$ at $2\le M\le 22$). Meanwhile, the table
shows that the NBD calculated slopes $\varphi_q$  yield the values
markedly distinguished from the observed ones in the stronger intermittency
case with possible non-thermal phase transition ($7\le M\le 17$, $10\le
M\le 17$), and trend to coincide each other at weak intermittency. In our
opinion, this is because of input two-particle character of $k$
calculated  from (\ref{nbd}): flattening of the NBD prediction with $q$
increase is visible at $q\ge 6$ (fig. 1).

Such a difficult meat when the NBD is applied to the intermittency search
makes   it problematic to describe the data. Let us note that invalidity
of the interpretation of multiparticle  production at high energies in the
NBD terms was also showed recently \cite{fos,dm,du}.

The gaussian (``log-normal'') approximation (GA)
predicts the relations for the slopes $\varphi_q$ to be defined
as \cite{bp1}

\begin{equation}
\varphi_q = \frac{\varphi_2}{2}\, q\, (q-1).
\label{gs}
\end{equation}

In  figs. 2 and 3 we show the GA predictions for the $d_q$ (\ref{d}) and
$\lambda_q$ (\ref{l}) using the $\varphi_2$-indexes at  $2\le M\le 28$
only. This is done by force of almost equal quantities of the $d_2$ for all
the cases of the $M$-intervals considered. One can conclude the GA seems
to be also hard to describe the intermittency effect especially when the
phase transition is possible.
In this sense it is interesting to note that the GA and the NBD approach
both are found \cite{eb}  to describe the cascade with small number of
steps, reflecting invalidity of these approximations for the multiplicity
asymptotics. The non-gaussian character of the correlations in multihadron
production was also shown earlier \cite{dw,d}.

The same result though less clearly  is seen (fig. 2 and 3) if the
simple scale-invariant cascade model proposed by Ochs and
Wosiek \cite{ow} is applied. In this case a modified power-law,

\begin{equation}
\langle F_q(\delta \eta )\rangle \propto
[g(\delta \eta)]^{\phi_q}\, ,
\label{ow}
\end{equation}
was showed to be universal dependence for large class of models and
of the available data of the one- and multidimensional SFM
\cite{o}. Here $g(\delta \eta )$ is the model-dependent function
of $\langle F_2\rangle $, so that $\phi_q = r_q\phi_2$.
The conclusion agrees with the earlier observations \cite{pl3} and shouts to
one another obtained in hadronic interactions \cite{22z}.

Thus all used model approximations,  being based on the second order
moments, indicate {\it multiparticle character} of the dynamical
fluctuations at (possible) non-thermal ({\it peak}) phase transition.

\section{Conclusions}

%To recapitulate,
The study of the intermittency phenomenon up to higher
ranks of the scaled factorial moments of the pseudorapidity distributions
of charged particles produced in central C--Cu interactions at 4.5 GeV/$c$
per nucleon is performed. To eliminate the problem of the spectrum shape
and to reach the higher multiple fluctuations the transformed variables
are used. The unbiased modification is applied to the moments
normalization to avoid a bias at small bins. The study shows existence of
intermittent-like increase of the moments at all the orders considered,
leading to a pronounced multifractality. The higher moment analysis allows
to reveal two sharply differed increases with bin
size, one of which    indicates stronger intermittency and then
non-thermal phase formation, and another one is usually connected with
the weak intermittency when the critical $q$-order is not reached. While
this comes from the fitting procedure at different bin intervals we are
inclined to consider the fact obtained as  a manifestation of
different regimes of particle production at different scales of random
cascading with a non-thermal phase transition inside. The results are
compared to ones given in the assumption of a validity of the
 negative binomial
distribution,  gaussian approximation
and scale-invariant mass-splitting branching model. The essential
multiparticle character of the phase transition is indicated. Accounting
possible hadronization influence on the dynamical correlations
\cite{bpla} and search
for the manifestations of a new matter formation the investigation of
higher multiparticle fluctuations in nuclear collisions at intermediate
energies gives evidence of further study of the effects observed.

\vspace{0.5cm}

{\large \bf Acknowledgements}\\

E.S., L.G. and G.L. thank J. Manjavidze and E. Gurvich for useful
discussions
and L. Rurua and B. Tarkhnishvili for the help in the paper preparation. It
is a pleasure to thank Professors K. Fia{\l}kowski, P.L. Jain, W. Kittel, I.
Otterlund, N. Schmitz  and R.M Weiner for sending us the papers not
available by the known causes.

{\scriptsize

%\newpage
\pagestyle{empty}

%\end{document}
}

\bigskip
{\large \bf Figure captions }
\medskip

{\bf Fig. 1.} The log-log plots of the modified scaled factorial moments
(\ref{fb}) vs. the number of divisions. The curves present the NBD
calculations (\ref{nbd}) and the straight lines show the least-squares
fits at $10\le M\le 17$.
\medskip

{\bf Fig. 2.} The codimensions $d_q$ versus $q$-order.
\medskip

{\bf Fig. 3.} The $\lambda _q$-functions (\ref{l}).

\vspace{0.6cm}
\begin{flushleft}
{\bf Table 1}
\end{flushleft}
\smallskip
The intermittency indexes $\varphi_q$ compared to the NBD predictions.
The errors present the covariance matrix estimators of the linear
least-squares fits.

\smallskip
\begin{flushleft}
\begin{tabular}{|c|c|c|c|c|} \hline  \hline
$q$&$2\le M\le 42$&$2\le M\le 28$&$2\le M\le 22$&$4\le M\le 15$ \\ \hline
2&0.017$\pm $0.001~~~~~~~~-- &0.011$\pm $0.001~~~~~~~~-- &0.009$\pm
    $0.001~~~~~~~~-- &0.009$\pm $0.001~~~~~~~~--\\
3&0.058$\pm $0.001~~~0.059 &0.039$\pm $0.001~~~0.035 &0.037$\pm
   $0.002~~~0.029 &0.032$\pm $0.004~~~0.030\\
4&0.130$\pm $0.002~~~0.113 &0.087$\pm $0.003~~~0.066 &0.092$\pm
   $0.004~~~0.029 &0.079$\pm $0.009~~~0.057\\
5&0.121$\pm $0.004~~~0.180 &0.124$\pm $0.007~~~0.106 &0.164$\pm
   $0.008~~~0.088 &0.153$\pm $0.019~~~0.092\\
6& -- &0.113$\pm $0.015~~~0.153 &0.233$\pm $0.013~~~0.127 &0.234$\pm
   $0.035~~~0.133\\
7& -- & -- &0.268$\pm $0.021~~~0.172 &0.275$\pm $0.059~~~0.180\\
8& -- & -- &0.207$\pm $0.031~~~0.222 &0.207$\pm $0.091~~~0.231\\
\hline
\end{tabular}
\end{flushleft}

\smallskip
\begin{flushleft}
\begin{tabular}{|c|c|c|c|} \hline  \hline
$q$&$10\le M\le 22$&$7\le M\le 17$&$10\le M\le 17$\\ \hline
2&0.004$\pm $0.002~~~~~~~~--
    &0.012$\pm $0.002~~~~~~~~-- &$-0.013\pm $0.004~~~~~~~~~-- \\
3&0.051$\pm $0.009~~~0.020
    &0.054$\pm $0.006~~~0.035 &0.020$\pm $0.018~~~0.00003 \\
4&0.194$\pm $0.030~~~0.039
    &0.178$\pm $0.015~~~0.069 &0.149$\pm $0.046~~~0.0019 \\
5&0.450$\pm $0.044~~~0.062
    &0.432$\pm $0.031~~~0.110 &0.489$\pm $0.100~~~0.0035 \\
6&0.894$\pm $0.076~~~0.089 &0.837$\pm $0.058~~~0.160
    &1.187$\pm $0.178~~~0.0055 \\
7&1.416$\pm $
    0.121~~~0.119 &1.346$\pm $0.101~~~0.216 &2.239$\pm $0.277~~~0.0073 \\
8&1.643$\pm $
    0.200~~~0.153 &1.847$\pm $0.171~~~0.278 &3.490$\pm $0.401~~~0.0098 \\
\hline
\end{tabular}
\end{flushleft}

\end{document}